\documentstyle[12pt,prd,aps,epsfig,preprint]{revtex}
\begin{document}
\draft
\title{Collisional Effects in Isovector Response Function of Nuclear
       Matter at Finite Temperature}
\author{
S.\  Ayik$^{1}$,  A. Gokalp$^{2}$, O. Yilmaz$^{2}$, K.
Bozkurt$^{2}$   }
\address{$^{1}$Physics Department, Tennessee Technological University,
Cookeville, TN 38505, USA}
\address{$^{2}$Physics Department, Middle East Technical University,
06531 Ankara, Turkey}
\date{\today}
\maketitle
\begin{abstract}
The dipole response function of nuclear matter at zero and finite
temperatures is investigated by employing the linearized version
of the extended TDHF theory with a non-Markovian binary collision
term. Calculations are carried out for nuclear dipole vibrations
by employing the Steinwedel-Jensen model and compared with
experimental results for $^{120}Sn$ and $^{208}Pb$.
\end{abstract}

~~~~~~~~

\narrowtext

Giant resonances, in particular giant dipole resonances (GDR), in
medium-weight and heavy nuclei have been the subjects of extensive
experimental and theoretical studies during the last several years
\cite{R1}. A large amount of experimental information is now
available concerning the properties of GDR built on the ground and
the excited states of the nuclei revealing the properties of the
collective motion of nuclear many-body systems at zero and finite
temperatures. The mean resonance energy is observed not to change
much with the excitation  energy, or the temperature, but the
recent experimental investigations show that the  width of the
resonance becomes broader as excitation energy increases with a
possible saturation at high temperatures. This temperature
dependence of the GDR width is still one of the open problems in
the studies of nuclear collective response and its damping
mechanisms at zero and finite temperatures \cite{R2,R3,R4}.

The theoretical investigations of the nuclear collective response
employing the random phase approximation (RPA) theory have been
quite successful in describing the mean resonance energies
\cite{R5}. However, the RPA theory is not
suitable for describing the damping of collective excitations and
therefore investigations based on the RPA theory have not been
able to explain the increase of the width of GDR with temperature
\cite{R6} .

There are different mechanisms involved in the damping of the
nuclear collective state. A part of the damping is due to the
coupling of the collective mode to external degrees of freedom
resulting in the cooling of the system by particle emission giving
rise to escape width. Furthermore, the collective mode also
acquires an intrinsic width as a consequence of its coupling to
the internal degrees of freedom. The resulting spreading width
which is thus due to the mixing of the collective mode with more
complicated doorway states makes essentially the large part of the
contribution to damping in medium-weight and heavy nuclei. There
are essentially three different theoretical approaches for the
calculation of the spreading widths. In the first case, the
temperature dependence of the width is explained by the coherent
mechanism due to adiabatic coupling of the giant resonance with
thermal surface deformations \cite{R7,R8}, which is particularly
important at low temperature. In the second approach, the
mechanism of damping is due to the coupling with incoherent two
particle-two hole (2p-2h) states resulting from the increasing
rate of collisions between the nucleons with temperature which is
usually referred to as the collisional damping \cite{R9,R10}. The
collisional damping is relatively weak at low temperature, but its
magnitude becomes large with increasing temperature. The last
mechanism is the Landau damping which is due to the spreading of
the collective mode on non-collective particle-hole (p-h)
excitations. Most investigations of nuclear response that have
been carried out so far are based on the coherent damping or the
collisional damping mechanisms \cite{R11,R12}.

In this work, we perform a linear response treatment of the
nuclear collective mode by including the collisional damping. The
small amplitude limit of the extended the time-dependent
Hartree-Fock (TDHF) theory, in which the collisional damping due
to the incoherent 2p-2h decay is included in the form of a
non-Markovian collision term, provides an appropriate framework
for investigating the damping widths of collective modes at zero
and finite temperatures \cite{R10,R11,R12,R13}. We employ the
extended TDHF theory to study the isovector response function with
collisional effects of nuclear matter at zero and finite
temperatures in semiclassical approximation using a simplified
effective Skyrme force.

The equation of motion of the single particle density matrix
$\rho(t)$ in the extended TDHF approximation is given by a
transport equation \cite{R13}
\begin{equation}\label{e1}
i\hbar \frac{\partial}{\partial t}\rho-[h(\rho),\rho]=K(\rho)~~,
\end{equation}
where $h(\rho)$ is an effective mean-field Hamiltonian and the
right-hand-side represents a non-Markovian collision term, which
can be expressed in terms of the correlated part of the
two-particle density matrix as $K(\rho)={Tr}_2[v,C_{12}]$ with the
effective residual interaction $v$. The correlated part of the
two-particle density matrix
$C_{12}=\rho_{12}-\widetilde{\rho_1\rho_2}$ where
$\widetilde{\rho_1\rho_2}$ represents the antisymmetrized product
of the single-particle density matrices, is given by the second
equation of the BBGKY hierarchy. In the extended TDHF theory, the
hierarchy is truncated at the second level by retaining only the
lowest-order terms in the residual interactions, thus neglecting
three-body correlations. Hence, the correlated part of the
two-particle density matrix $C_{12}$ satisfies the equation
\begin{equation}\label{e2}
i\hbar \frac{\partial}{\partial t}C_{12}-[h(\rho),C_{12}]=F_{12}
\end{equation}
where the source term is given by
\begin{equation}\label{e3}
F_{12}(\rho)=(1-\rho_{1})(1-\rho_{2})~v~\widetilde{\rho_{1}\rho_{2}}-
\widetilde{\rho_{1}\rho_{2}}~v~(1-\rho_{1})(1-\rho_{2})~~.
 \end{equation}

In order to study the isovector collective response of the system,
we include an external perturbation $F(\vec{r},t)$ into the
equation of motion,
\begin{equation}\label{e4}
  F(\vec{r},t)=\tau_3F(\vec{r})\left[e^{-iwt}+e^{iwt}\right]~~,
\end{equation}
where $\tau_3$ is the third component of the isospin operator and
the frequency of the one-body harmonic perturbation operator
contains a small imaginary part with the prescription
$\omega\rightarrow\omega+i\eta$ in accordance with the adiabatic
hypothesis. We obtain a description for small density fluctuations
$\delta\rho(t)=\rho(t)-\rho_0$ in linear response treatment by
linearizing the extended TDHF theory around a finite temperature
equilibrium state density $\rho_0$, and this way we obtain
\begin{equation}\label{e5}
i\hbar \frac{\partial}{\partial t}\delta\rho-[h_0,\delta\rho]-
[\delta h+F(\vec{r},t),\rho_{0}]=Tr_{2}[v,\delta C_{12}]
\end{equation}
where $\delta h=\left(\partial U/\partial \rho \right)_{0}
\delta\rho$ represents  small deviations in the effective
mean-field potential. Moreover, the small deviation of the
two-body correlations $\delta C_{12}(t)$ satisfies
\begin{equation}\label{e6}
i\hbar \frac{\partial}{\partial t}\delta C_{12}- [\delta
h+F(\vec{r},t),C_{12}^{0}]-[h_{0},\delta C_{12}]=\delta F_{12}~~.
\end{equation}

We look for a solution of Eq. 5 and Eq. 6 of the form
$\delta\rho(t)=\delta\rho(\omega)e^{-iwt}+h.c.$ where now the
small density fluctuations are given in terms of the proton and
neutron density matrices as $\delta\rho(t)=\rho_p(t)-\rho_n(t)$.
We note that in the small amplitude limit, the collision term is
also harmonic, $\delta K=tr_2[v,\delta C_{12}]=\delta
K(\omega)e^{-iwt}+h.c.$, and in momentum representation we then
obtain
\begin{eqnarray}\label{e7}
&&\left[ \hbar\omega+\epsilon(\vec{p}-\frac{\vec{k}}{2})
-\epsilon(\vec{p}+\frac{\vec{k}}{2})\right]
<\vec{p}+\frac{\vec{k}}{2}\mid\delta\rho(w)\mid
\vec{p}-\frac{\vec{k}}{2}>
-\left[f(\vec{p}-\frac{\vec{k}}{2})-f(\vec{p}+\frac{\vec{k}}{2})\right]
 \nonumber \\
&&\times\left\{<\vec{p}+\frac{\vec{k}}{2}\mid \delta h
\mid\vec{p}-\frac{\vec{k}}{2}> +2<\vec{p}+\frac{\vec{k}}{2}\mid
F(\vec{r}) \mid\vec{p}-\frac{\vec{k}}{2}>\right\} =
<\vec{p}+\frac{\vec{k}}{2}\mid \delta K(\omega)
\mid\vec{p}-\frac{\vec{k}}{2}>,
\end{eqnarray}
where
$f(\vec{q})=1/\left\{1+e^{\beta[\epsilon(\vec{q})-\mu]}\right\}$
is the Fermi-Dirac occupation factor. For simplified Skyrme
interaction, the density fluctuation $\delta n(\vec{r},t)$ induces
local changes in the mean field potential, therefore
\begin{equation}\label{e8}
<\vec{p}+\frac{\vec{k}}{2}\mid \delta h\mid
\vec{p}-\frac{\vec{k}}{2}>=2V_0<\vec{p}+\frac{\vec{k}}{2}\mid
\delta\rho(w)\mid \vec{p}-\frac{\vec{k}}{2}>
\end{equation}
where in terms of local proton and neutron mean field potentials
$\delta h$ is expressed as $\delta h = U_p(\vec{r},t)-U_n(\vec{r},t)
=2V_0\delta n(\vec{r},t)$.
Moreover, since
\begin{equation}\label{e9}
\int\frac{d^3p}{(2\pi\hbar)^3}
<\vec{p}+\frac{\vec{k}}{2}\mid\delta\rho(w)\mid
\vec{p}-\frac{\vec{k}}{2}>=\delta n(\vec{k},w)
\end{equation}
and
\begin{equation}\label{e10}
<\vec{p}+\frac{\vec{k}}{2}\mid F(\vec{r})\mid
\vec{p}-\frac{\vec{k}}{2}>=F(\vec{k})
\end{equation}
we finally obtain
\begin{eqnarray}\label{e11}
\delta n(\vec{k},w)-\left[ V_0\delta
n(\vec{k},w)+F(\vec{k})\right]\Pi_1(\vec{k},w)= \left[ V_0\delta
n(\vec{k},w)+F(\vec{k})\right]\Pi_2(\vec{k},w)~~.
\end{eqnarray}
The function $\Pi_1(\vec{k},w)$, which  is known as the
unperturbed Lindhard function, is given by
\begin{equation}\label{e12}
\Pi_1(\vec{k},w)=\frac{2}{(2\pi\hbar)^3}\int d^3p~
\frac{f(\vec{p}-\frac{\vec{k}}{2})-f(\vec{p}+\frac{\vec{k}}{2})}
{\hbar w-\epsilon(\vec{p}+\frac{\vec{k}}{2})
+\epsilon(\vec{p}-\frac{\vec{k}}{2})+i\eta}
\end{equation}
and the function  $\Pi_2(\vec{k},w)$ is obtained from
\begin{equation}\label{e13}
[V_0\delta n(\vec{k},w)+F(\vec{k})]
\Pi_2(\vec{k},w)=\frac{2}{(2\pi\hbar)^3}\int~d^3p~
\frac{<\vec{p}+\frac{\vec{k}}{2}\mid \delta
K(w)\mid\vec{p}-\frac{\vec{k}}{2}>} {\hbar
w-\epsilon(\vec{p}+\frac{\vec{k}}{2})+\epsilon(\vec{p}-\frac{\vec{k}}{2})+i\eta}~~.
\end{equation}
The retarded response function which is defined by
\begin{equation}\label{e14}
\delta n(\vec{k},w)=\Pi_R(\vec{k},w)F(\vec{k})
\end{equation}
is then obtained as
\begin{equation}\label{e15}
\Pi_R(\vec{k},w)=\frac{\Pi_0(\vec{k},w)}{1-V_0\Pi_0(\vec{k},w)}
\end{equation}
with $\Pi_0(\vec{k},w)=\Pi_1(\vec{k},w)+\Pi_2(\vec{k},w)$.

From Eq. 6, it is possible to obtain a closed form expression for
small deviation of two-body correlations $\delta C_{12}$, that is
valid for collective vibrations (for details please refer to
\cite{R10}),
\begin{eqnarray}\label{e16}
\delta C_{12}(t)=
-\frac{i}{\hbar}\int^t~dt'\rho_1^0\rho_2^0e^{-ih_0(t-t')}[\delta
\Phi(t'),v] e^{ih_0(t-t')}(1-\rho_1^0)(1-\rho_2^0)+h.c.
\end{eqnarray}
where $\delta \Phi(t)$ is the distortion function associated with
the single-particle density matrix, and it is related to the small
vibrations in the single-particle density matrix $\delta\rho(t)$
according to $\delta\rho(t)=[\delta \Phi(t),\rho_0]$. We then
obtain the expression for the linearized collision term by
evaluating the matrix element $<\vec{p}+\frac{\vec{k}}{2}\mid
\delta K(\omega)\mid \vec{p}-\frac{\vec{k}}{2}>$ in which we
retain $\vec{k}$-dependence only in distortion function
\cite{R10}. Then, the collisional response function
$\Pi_2(\vec{k},w)$ can be expressed as
\begin{equation}\label{e17}
\Pi_2(\vec{k},w)=\frac{1}{(2\pi\hbar)^3}\int
d^3p_1d^3p_2d^3p_3d^3p_4\left(\frac{\Delta
Q}{2}\right)^2\frac{W(12;34)}{\pi}~
\frac{f_1f_2\overline{f}_3\overline{f}_4-
\overline{f}_1\overline{f}_2f_3f_4} {\hbar
w-\epsilon_3-\epsilon_4+\epsilon_1+\epsilon_2+i\eta}
\end{equation}
where $\overline{f}_i=1-f_i$, ~$\Delta Q=Q_1+Q_2-Q_3-Q_4$ with
$Q_i=1/\left[\hbar w-\epsilon(\vec{p_i}+\frac{\vec{k}}{2})
+\epsilon(\vec{p_i}-\frac{\vec{k}}{2})\right]$, and W(12;34)
denotes the basic two-body transition rate, which can be expressed
in terms of the spin averaged proton-neutron scattering cross
section as
\begin{equation}\label{e18}
W(12;34)=\frac{1}{(2\pi\hbar)^3}~\frac{4\hbar}{m^2}~
\left(\frac{d\sigma}{d\Omega}\right)_{pn}~
\delta^3(\vec{p}_1+\vec{p}_2-\vec{p}_3-\vec{p}_4)~~.
\end{equation}
The strength distribution function is obtained from the imaginary
part of the retarded response function \cite{R14}
\begin{equation}\label{e19}
S(\vec{k},w)=-\frac{1}{\pi}Im\Pi_R(\vec{k},w)~~.
\end{equation}

In our calculations, we employ a simplified Skyrme interaction
\begin{equation}\label{e20}
  v=t_0(1+x_0P_\sigma)\delta(\vec{r})
  +\frac{1}{6}t_3(1+x_3P_\sigma)\rho^\alpha(\vec{R})\delta(\vec{r})
\end{equation}
with $\vec{r}=\vec{r_1}-\vec{r_2}$ and
$\vec{R}=(\vec{r_1}+\vec{r_2})/2$. The local potential for protons
is then given by
\begin{eqnarray}\label{e21}
  U_p(\vec{r},t)=&&t_0\left(1+\frac{1}{2}x_0\right)\rho(\vec{r},t)-
  t_0\left(\frac{1}{2}+x_0\right)\rho_p(\vec{r},t)\nonumber \\
  &&+\frac{1}{12}t_3\rho^\alpha(\vec{r},t)\left[(2+\alpha)
  \left(1+\frac{1}{2}x_3\right)\rho_p(\vec{r},t)\right.\nonumber \\
  &&\left.-2\left(\frac{1}{2}+x_3\right)\rho(\vec{r},t)-
  \alpha\left(\frac{1}{2}+x_3\right)\frac{\rho_p^2(\vec{r},t)+
  \rho_n^2(\vec{r},t)}{\rho(\vec{r},t)}\right]
\end{eqnarray}
with a similar expression for neutrons. In linear response
approximation, the coupling constant $V_0$ for dipole vibrations
becomes
\begin{equation}\label{e22}
  V_0=-\frac{1}{2}t_0\left(\frac{1}{2}+x_0\right)-
\frac{1}{12}t_3\rho^\alpha_0\left(\frac{1}{2}+x_3\right)
\end{equation}
where $\rho_0$ is the saturation density of nuclear matter. In our
analysis we consider in particular the Skyrme SLy4 force with the
parameters \cite{R15} $t_0=-2488.91~MeV.fm^3$,
$t_3=13777~MeV.fm^{7/2}$, $x_0=0.834$, $x_3=1.354$ and
$\alpha=1/6$, which results for $V_0$ in the value $V_0=85~MeV$.

In order to apply our results to finite nuclei, we work within the
framework of Steinwedel and Jensen model for nuclear dipole
oscillations \cite{R6}. In this model neutrons and protons
oscillate inside a sphere of radius R given by the expression
\begin{equation}\label{e23}
\rho_p(\vec{r},t)-\rho_n(\vec{r},t)=F\sin
(\vec{k}\cdot\vec{r})e^{iwt}~~,
\end{equation}
the total density remaining equal to the saturation density
$\rho_0$ of nuclear matter and the wavenumber k is given by
$k=\pi/2R$. We apply Steinwedel and Jensen model to GDR in
$^{120}Sn$ and $^{208}Pb$, and we take $R=5.6$ fm $k=0.28~fm^{-1}$
for $^{120}Sn$ and $R=6.7$ fm $k=0.23~fm^{-1}$ for $^{208}Pb$
according to $R=1.13A^{1/3}$.

As a result of the approximate treatment, the collisional response
function $\Pi_2(\vec{k},\omega)$ has a singular behavior arising
from the pole of the distortion functions, $Q_i=1/\left[\hbar
w-\epsilon(\vec{p_i}+\frac{\vec{k}}{2})
+\epsilon(\vec{p_i}-\frac{\vec{k}}{2})\right]$. We avoid this
singular behavior by incorporating a pole approximation. In the
distortion functions, we make the replacement $\omega\rightarrow
\omega_D-i\Gamma/2$ where $\omega_D$ and $\Gamma$ are determined
from $1-V_0\Pi_1(\vec{k},\omega)=0$ at each temperature that is
considered. Furthermore, we neglect the real part $Re\Pi_2$ of the
function $\Pi_2$ in our calculations. In the calculations of the
collisional response function, using conservation laws and
symmetry properties, it is possible to reduce the twelve
dimensional integrals to five fold integrals by incorporating the
transformations into the total momenta
$\vec{P}=\vec{p}_1+\vec{p}_2$, $\vec{P'}=\vec{p}_3+\vec{p}_4$, and
relative momenta $\vec{q}=(\vec{p}_1-\vec{p}_2)/2$,
$\vec{q}~^\prime=(\vec{p}_3-\vec{p}_4)/2$ before and after the
collisions. The integral over $\vec{P'}$ can be performed
immediately. The delta function
$\delta(\hbar\omega-\epsilon'+\epsilon)$ in $Im\Pi_2(\vec{k},w)$
where $\epsilon=\vec{q}~^{2}/m$ and
$\epsilon'=\vec{q}~^\prime~^{2}/m$ are the energies of two
particle system in the center of mass frame before and after the
collision makes it possible the reduce the integrals further using
familiar methods from the Fermi liquid theory \cite{R17}. Then, we
evaluate the remaining five dimensional integrals numerically by
employing a fast algorithm. In the evaluation of momentum
integrals, we neglect  the angular anisotropy of the cross
sections and make the replacement
$\left(d\sigma/d\Omega\right)_{pn}\rightarrow\sigma_{pn}/4\pi$
with $\sigma_{pn}=40$ mb.

We show our results for the response function with and without the
collision term in Fig. 1 for $^{120}Sn$ and in Fig. 2 for
$^{208}Pb$ as a function of experimental temperature $T^*$ where
we also present the comparison with the normalized experimental
data \cite{R4}. The experimental temperature $T^*$ is related to
the temperature parameter in the Fermi-Dirac function $f(\epsilon,
T)$ as $T=T^*\sqrt{a_E/a_F}$, where $a_E$ denotes the energy
dependent empirical level density parameter and
$a_F=A\pi^2/4\epsilon_F$ denotes Fermi gas level density parameter
\cite{R4,R16}. In our calculations, we use the temperature values
T in the Fermi-Dirac function that are related to the experimental
temperatures in this manner.

We first note that, at the RPA level that is without the
contribution of the collision term in calculation of the nuclear
response, the position of the peak of the response functions
do not change with temperature. As a matter of fact, for
$^{208}Pb$ the peak is at $\omega=12.3~MeV$ for
$T^*=1.34,~1.62,~1.85,~2.05~~MeV$ while for $^{120}Sn$ it occurs
at $\omega=14.3~MeV$ for $T^*=1.36,~2.13~~MeV$ and at
$\omega=14.2~MeV$ for $T^*=2.67,~3.1~~MeV$. This behavior of the
peak energy with temperature is in accordance with the
experimental results where it is observed that the mean-energy of
the dipole response is almost constant for $^{208}Pb$ when $T^*$
changes between 1.3 to 2.0 MeV while a decrease of 1.5 MeV is
observed in $^{120}Sn$ when $T^*$ changes from 1.2 to 3.1 MeV
\cite{R4}. However, the average positions of the peak values of
the strength functions are slightly below the
experimental values, which are $\omega=15.4~MeV$ for $^{120}Sn$ and
$\omega=13.4~MeV$ for $^{208}Pb$. This discrepancy may be due to
the nature of the effective Skyrme force that we employ. Moreover,
the value for $k=\pi/2R$ that is used in Steinwedel and Jensen
model depends on the value of $R_0$ used in $R=R_0A^{1/3}$, and
changing k somewhat also produces a change in the position of the
peak but in general the above conclusions are not affected. Furthermore,
These results for the peak position of the strength functions
are in accordance with the earlier RPA calculations.

Dotted lines in figures 1-2 show the strength functions including
collisional damping mechanism. Since, we neglect the real part of
the collisional response the peak values of the strength do not change,
but the collisional mechanism introduces a spread, in particular at the
high frequency side of the strength functions and this spread becomes
more pronounced with increasing temperature.
In order to illustrate the effect of the collision term more
drastically, in figure 3, we show the response function with and without the
collision term for $^{120}Sn$ and for $^{208}Pb$ at temperatures T=2,4 MeV.

Rather simple description presented in this paper is able to
explain certain aspects of giant dipole excitations in $^{120}Sn$
and  $^{208}Pb$, but do not produce a good description of the
experimental strength functions as a function of temperature. One
important element missing in the calculations is the coherent
damping mechanism due to coupling dipole vibrations with low
frequency collective surface modes \cite{R18,R19,R20}. This
mechanism is especially important for describing the details of
the strength distributions at low temperature. In the continuation
of this work, we plan to improve this simple description by
incorporating the coherent damping mechanism in Thomas-Fermi
approximation.

\newpage
\begin{figure}
\epsfig{figure=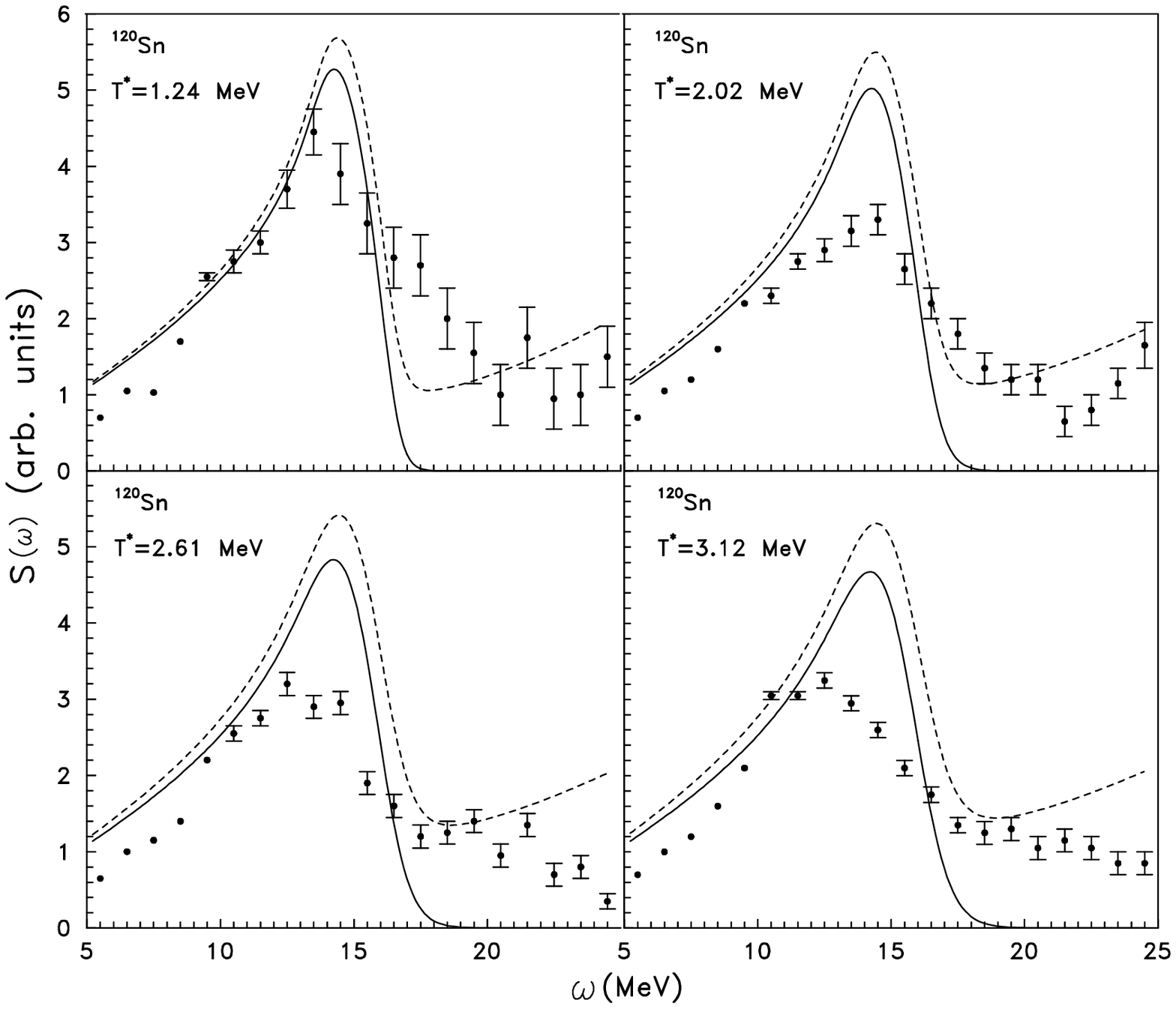,height=20cm, angle=0}\vspace*{-3.0cm}
\caption{The GDR strength function of $^{120}$Sn. Solid and dashed
lines show the response function without and with the collision
term, respectively. The normalized data is taken from [4].}
\end{figure}

\begin{figure}
\vspace*{1.0cm} \epsfig{figure=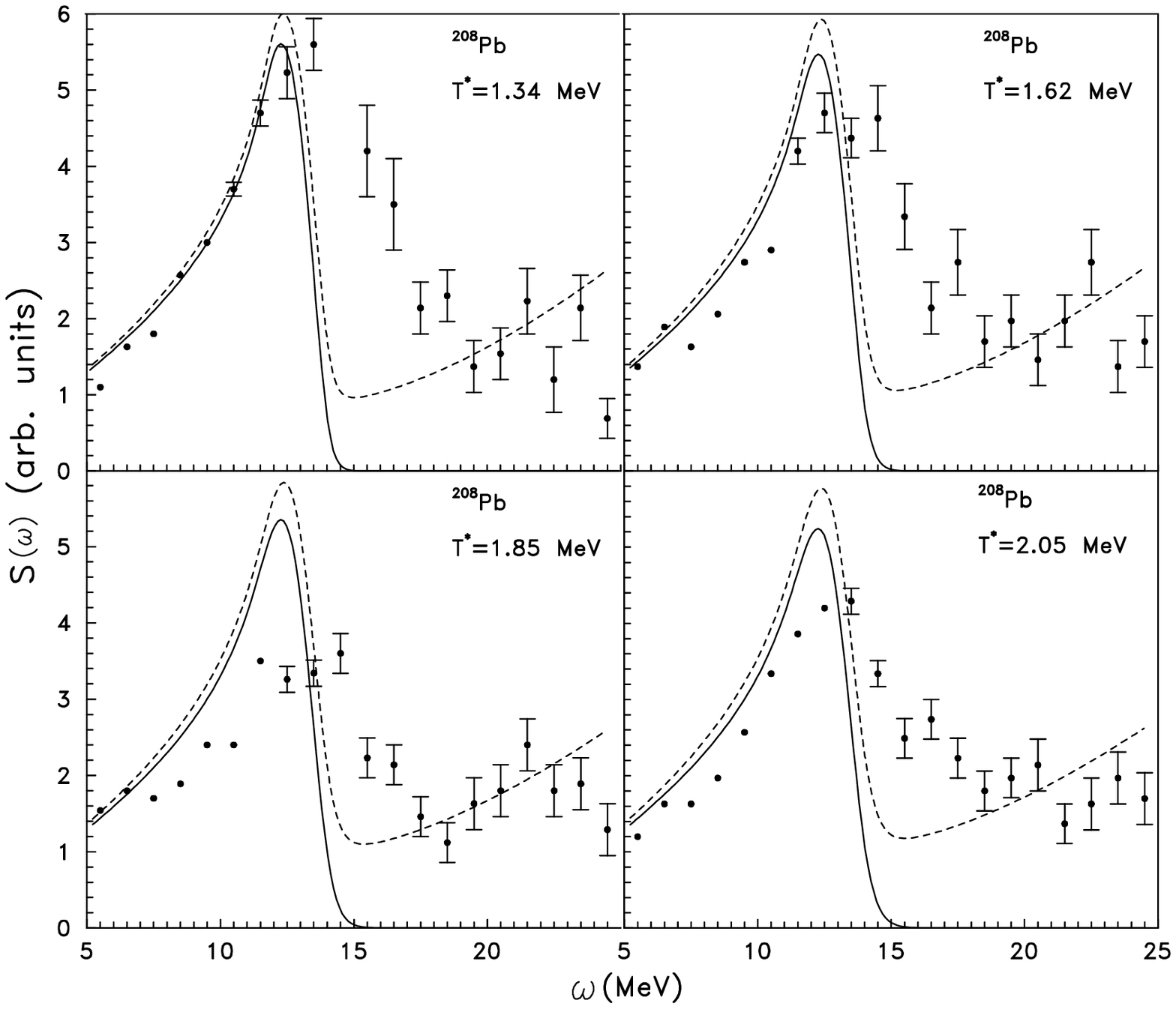,height=20cm,
angle=0}\vspace*{-3.0cm} \caption{The GDR strength function of
$^{208}$Pb. Solid and dashed lines show the response function
without and with the collision term, respectively. The normalized
data is taken from [4].}
\end{figure}

\newpage

\begin{figure}
\epsfig{figure=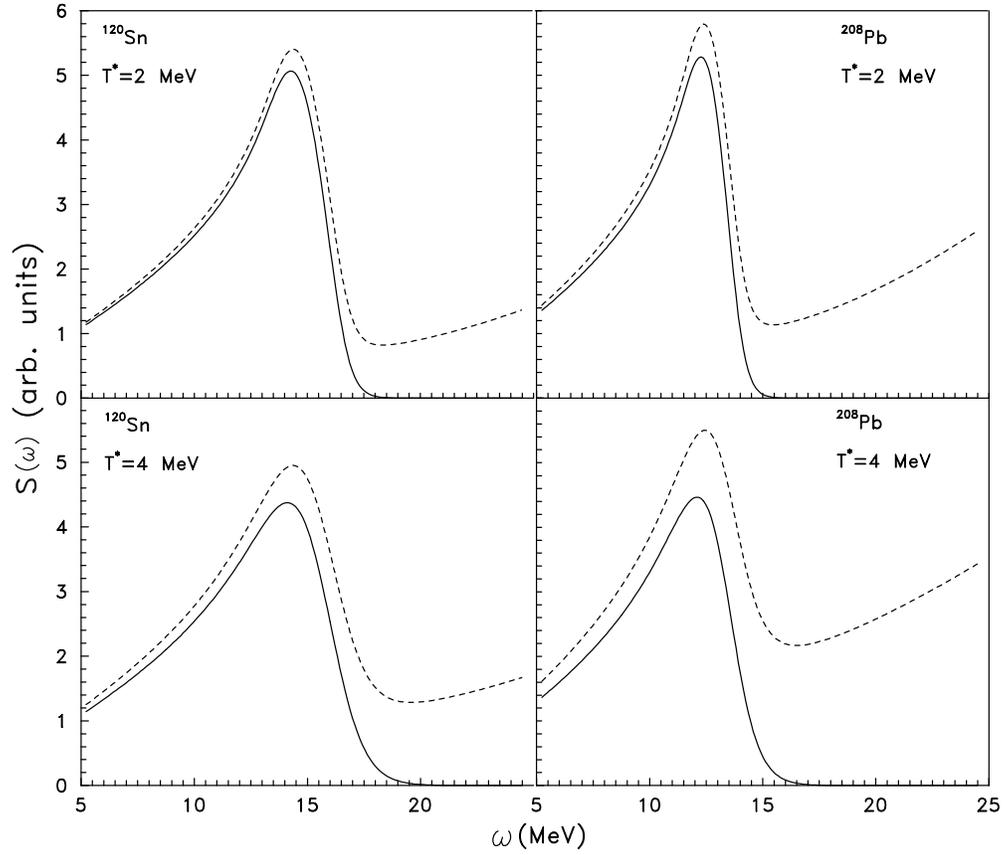,height=20cm}\vspace*{-3.0cm}
\caption{The GDR strength function of $^{120}$Sn and $^{208}$Pb at
$T^*=2,~4$ MeV. Solid and dashed lines show the response function
without and with the collision term, respectively.}
\end{figure}

\end{document}